# Intrinsic auxeticity and mechanical anisotropy of $Si_9C_{15}$ siligraphene

Jianli Zhou[∥], Jian Li[∥], Jin Zhang[*]

School of Science, Harbin Institute of Technology, Shenzhen 518055, PR China

[∥]Contributed equally.
[*]Corresponding author. E-mail address: jinzhang@hit.edu.cn (J. Zhang).

**Abstract:** The graphene-like two-dimensional (2D) silicon carbide or siligraphene has attracted remarkable attentions, owing to its fascinating physical properties. Nevertheless, the first high-quality siligraphene, i.e., monolayer $Si_9C_{15}$ was synthesised very recently, which exhibits an excellent semiconducting behaviour. In this work, we investigate the mechanical properties of $Si_9C_{15}$ siligraphene by using atomistic simulations including density functional theory (DFT) calculations and molecular dynamics (MD) simulations. Both methods confirm the existence of intrinsic negative Poisson's ratios in $Si_9C_{15}$ siligraphene, which, as illustrated by MD simulations, is attributed to the tension-induced de-wrinkling behaviours of its intrinsic rippled configuration. Different de-wrinkling behaviours are observed in different directions of $Si_9C_{15}$ siligraphene, which result in the anisotropy of its auxetic property. The fracture property of $Si_9C_{15}$ siligraphene is similarly anisotropic, but relatively large fracture strains are observed in different orientations, indicating the stretchability of $Si_9C_{15}$ siligraphene. The stretchability together with the strain-sensitive bandgap of $Si_9C_{15}$ siligraphene observed in DFT calculations indicates the effectiveness of strain engineering in modulating its electronic property. The combination of unique auxetic property, excellent mechanical property and tunable electronic property may render $Si_9C_{15}$ siligraphene a novel 2D material with multifunctional applications.

# 1. Introduction

The successful mechanical exfoliation of monolayer graphene flakes from graphite in 2004 [1] has prompted the incredible research interests in two-dimensional (2D) materials. Since then, many new 2D materials such as transition metal-dichalcogenide [2], silicene [3], hexagonal boron nitride (hBN) [4], and black phosphorus [5] have been synthesized by the aforementioned mechanical exfoliation method and some other synthesis methods such as electrochemical intercalation, ball milling, electrochemical intercalation, chemical vapor deposition, physical vapor deposition, to name a few [6, 7]. Compared to their bulk counterparts, 2D materials possess many superior material properties, such as significant quantum effects [8], extremely high surface area-to-volume ratio [9], extra-large electrical and thermal conductivities [10, 11], and extraordinary mechanical strength [12]. Inspired by the remarkable properties discovered in 2D materials as well as the development of synthesis technique at the nanoscale, many efforts recently have been made to explore the possible existence of some other 2D materials [13, 14].

As a semiconductor with the widest range of energy bandgap, silicon carbide (SiC) has attracted extensive research interests. Specifically, the wide-bandgap semiconducting property together with the high thermal capability of SiC makes it become a key material in various technological applications including high-power electronics, photonic devices, high-temperature devices and quantum information processing [15-17]. Inspired by the unique property, especially outstanding semiconducting properties, the exploration of 2D form of SiC is currently receiving the most attention from the community of 2D materials. It is expected that 2D SiC can not only retain the overall SiC properties but also exhibit unique optical and electronic properties due to the reduced dimensionality and quantum confinement [18, 19]. As a binary compound of carbon and silicon, 2D SiC can exist in a variety of compositions and hence structures i.e., $Si_xC_y$, which include the siligraphene possessing the similar honeycomb lattice of graphene and the non-siligraphene crystals with structures beyond the hexagonal lattice. To date, many 2D SiC materials have been theoretically predicted to be structurally stable. For example, C-rich siligraphene ($SiC_2$, $SiC_3$, $SiC_5$ and $SiC_7$) [20-27] and Si-rich siligraphene ($Si_3C$, $Si_5C$ and $Si_7C$) have been proposed by high-throughput density functional theory (DFT) calculations [24-27]. Recently, DFT calculations were also employed to seek the non-siligraphene crystals such as penta-$SiC_2$, $SiC_6$, $\alpha$-$SiC_7$, $SiC_8$, to name a few [28-32]. Meanwhile, many theoretical efforts also have been made to reveal the material properties of 2D SiC materials. Through atomistic simulations including DFT calculations and molecular dynamics (MD) simulations, many superior properties and thus novel potential applications of 2D SiC have been observed [29, 33-36]. Although a great progress has been made in theoretically predicting structures of 2D SiC materials, few of these predicted 2D SiC materials have been synthesized in the laboratory due to the lack of the corresponding experimental synthetic routes. Very recently, Gao et al. [37] made a big progress in the synthesis of 2D SiC by successfully fabricating the high-quality $Si_9C_{15}$ siligraphene with the large-scale atomic monolayer. The possible exfoliation of freestanding monolayer $Si_9C_{15}$ from metal substrates together with its good air stability and semiconducting behaviour makes $Si_9C_{15}$ siligraphene appealing for applications in 2D electronics and photonics. In these applications, the monolayer $Si_9C_{15}$ is required to retain its structural integrity under the possible external stimuli. Besides, because the strain engineering is usually an efficient method to modify physical properties of 2D materials [38-41], the mechanics of $Si_9C_{15}$ siligraphene is also important for proper functioning of $Si_9C_{15}$ siligraphene-based devices. Thus, to promote the future applications of newly synthesized $Si_9C_{15}$ siligraphene, it becomes necessary to achieve a comprehensive understanding of its mechanical properties.

In this paper, the mechanical properties of $Si_9C_{15}$ siligraphene are comprehensively investigated by DFT calculations and MD simulations. The mechanical and thermodynamic stabilities of freestanding $Si_9C_{15}$ siligraphene are also verified by the Born stability criteria and *ab initio* molecular dynamics (AIMD) simulations, respectively. Both DFT calculations and MD simulations illustrate a stable rippled configuration intrinsically existing in $Si_9C_{15}$ siligraphene, the de-wrinkling of which under uniaxial tension results in its auxetic behaviours. The auxetic behaviours as well as the fracture behaviours of $Si_9C_{15}$ siligraphene are found to be dependent on the loading direction. Finally, the temperature effect on auxetic and fracture properties of $Si_9C_{15}$ siligraphene is also examined.

## 2. Computational methodology

### 2.1 DFT calculations

DFT calculations were conducted to majorly examine the fundamental elastic properties and the thermodynamic stability of freestanding monolayer $Si_9C_{15}$. Here, all DFT calculations were conducted through the Vienna ab initio simulation package (VASP) [42] with the plane-wave cutoff energy in the Perdew-Burke-Ernzerhof (PBE) generalized gradient approximation (GGA) [43] being set as 400 eV for the exchange correlation potential. The convergence criterion for the electronic self consistence-loop was set as $10^{-5}$ eV, while the force threshold of structural optimization was set as $10^{-2}$ eV/Å. The k-point mesh with a density of 0.25/Å was sampled in the in-plane Brillouin zone of the unit cell. Periodic boundary conditions were applied along all three directions to exclude the boundary effect. Moreover, a vacuum layer of 25 Å was added in the thickness direction to avoid image-image interactions. After obtaining the energy minimized structures of monolayer $Si_9C_{15}$, its electronic properties were evaluated by using a k-point grid of $21 \times 21 \times 1$. In addition, AIMD simulations at the temperature of 500 K were also performed to verify the thermodynamic stability of monolayer $Si_9C_{15}$ by reducing the k-point grid into $2 \times 2 \times 1$. Here, the optimized structures of monolayer $Si_9C_{15}$ were simulated within the NVT (constant number of particles, volume and temperature) ensemble using Andersen thermostat at 500 K and a time step of 1 fs.

**2.2 MD simulations**

MD simulations were conducted to study the mechanical properties especially the fracture properties of $Si_9C_{15}$ siligraphene, since dynamics factors such as the temperature effect can be considered in MD. Moreover, MD simulations have the capability to study a $Si_9C_{15}$ siligraphene with larger number of atoms. Here, a $Si_9C_{15}$ siligraphene sample with the size of 10 nm × 10 nm was considered in MD simulations. Periodic boundary conditions were applied in both in-plane directions, while a vacuum spacing of 5 nm was placed in the thickness direction. All MD simulations were conducted with the aid of open-source code LAMMPS [44] with the optimized Tersoff potential [45], which has been successfully employed in the previous MD simulations to describe 2D SiC materials [35, 36]. The velocity Verlet algorithm with a time step of 0.5 fs was employed to integrate equations of motion. Before conducting the mechanical study, conjugate gradient methods were used to obtain the equilibrium states of $Si_9C_{15}$ siligraphene considered here. In doing this, NPT (constant number of particles, pressure and temperature) simulations were conducted for 50 ps, in which the Nosé-Hoover thermostat and barostat were employed at a certain temperature and zero pressure. The mechanical behaviours of $Si_9C_{15}$ siligraphene were studied by gradually elongating the simulation box along one in-plane direction with a relatively low strain rate of 0.0005/ps. The strain rate selected here can be used to equivalently simulate the quasi-static loading. During the elongation process, the strain was taken as the relative change of the length of siligraphene, while the stress was calculated as the arithmetic mean of local stresses on all atoms [46].

## 3. Results and discussion

### 3.1 Mechanical properties at the ground state from DFT

As shown in Fig. 1a, $Si_9C_{15}$ siligraphene generally has a honeycomb lattice similar to that of graphene. Thus, there exist two principle directions in $Si_9C_{15}$ siligraphene, i.e., armchair (AC) and zigzag (ZZ) directions. In $Si_9C_{15}$ siligraphene, each C-C hexagon is surrounded by twelve Si-C hexagons. Because C-C and Si-C bonds have very different bond properties such as different bond lengths, the freestanding $Si_9C_{15}$ siligraphene is found to exhibit a rippled configuration with an amplitude around 1 Å after the structural optimization as shown in Fig. 1b. This intrinsic nonplanar configuration of $Si_9C_{15}$ siligraphene is totally in contrast to the planar structure of most other 2D materials, such as graphene, black phosphorus and hBN. To provide useful insights into the atomic bonding nature of $Si_9C_{15}$ siligraphene, we show in Fig. 1c its electron localization function (ELF), which is a spatial function having the ability to quantitatively distinguish different types of chemical bonds. The value of ELF ranges between 0 and 1. Specifically, an ELF close to 1 indicates strong covalent interactions or lone pair electrons, while a lower ELF corresponds to metallic or ionic bonds. As shown in Fig. 1c, values of ELF around the centre of each bond are more than 0.8, which indicates that atoms in $Si_9C_{15}$ siligraphene are covalently bonded. Moreover, the electron localization around the centre of Si-C bonds is broader than that of their C-C counterparts, which indicates the weaker stiffness of Si-C bonds. Thus, it is reasonable to expect that $Si_9C_{15}$ siligraphene should possess a weaker stiffness compared to graphene, which will be verified later in the following DFT and MD studies. Fig. 1d shows that the energy of $Si_9C_{15}$ siligraphene reaches equilibrium rapidly at the beginning of AIMD simulations and remains almost unchanged afterwards. At the end of AIMD simulations, no bond breakages or disorders are found in the structure. In other words, no topological change is observed in $Si_9C_{15}$ siligraphene during the entire AIMD simulation process, in spite of a larger out-of-plane deformation observed due to the thermal fluctuation at a finite temperature. These results indicate the thermodynamic stability of the freestanding $Si_9C_{15}$ siligraphene.

Based on the finite distortion method in DFT calculations, the linear elastic constants of $Si_9C_{15}$ siligraphene were calculated after assuming the thickness of $Si_9C_{15}$ siligraphene as 3.5 Å [34, 35]. As for a 2D material, there are four independent elastic constants, i.e., $C_{11}$, $C_{22}$, $C_{12}$ and $C_{66}$ in $Si_9C_{15}$ siligraphene. Values of these elastic constants obtained from DFT calculations are listed in the Supplementary Material. These elastic constants satisfy the Born criteria of $C_{11}C_{22}-C_{12}^2>0$ and $C_{11}$, $C_{22}$, $C_{66}>0$ [47], which further indicates the mechanical stability of $Si_9C_{15}$ siligraphene. Furthermore, these elastic constants can be utilized to evaluate the dependence of Young's modulus $E$, shear modulus $G$ and Poisson's ratio $v$ on the orientation $\theta$ of $Si_9C_{15}$ siligraphene by using the following equations [48]:

$$E(\theta)=1/\left[S_{11}a^4+S_{22}b^4+(S_{66}+2S_{12})a^2b^2\right],$$

$$G(\theta)=1/\left[4(S_{11}+S_{22}-2S_{12})a^2b^2+S_{66}(a^2-b^2)^2\right], \qquad (1)$$

$$v(\theta)=-E(\theta)/\left[(S_{11}+S_{22}-S_{66})a^2b^2+S_{12}(a^4+b^4)\right],$$

where $a = \cos(\theta)$ and $b = \sin(\theta)$. In Eq. 1, $S_{11}$, $S_{22}$, $S_{12}$ and $S_{66}$ are compliance constants related to the elastic constants [48].

Almost isotropic Young's modulus $E$ and shear modulus $G$ are found in the intrinsically nonplanar $Si_9C_{15}$ siligraphene (see Fig. 2a and 2b). For example, when the orientation $\theta$ shifts from $0^o$ (the ZZ direction) to $45^o$, $E$ decreases from 272 GPa to 260 GPa. In this process, $G$ is found to increase from 141 GPa to 150 GPa. As $\theta$ changes to $90^o$ (the AC direction), $E$ slightly grows to 261 GPa, while $G$ declines to 141 GPa. In other words, $E$ and $G$ in all orientations of $Si_9C_{15}$ siligraphenes fluctuate around mean values of 266 GPa and 146 GPa, respectively. The deviation of the average value of $E$ from the maximum or minimum $E$ is less than 2%, while this deviation is less than 4% for $G$. The intrinsic $E$ and $G$ of the present $Si_9C_{15}$ siligraphene are found to be smaller than values of graphene and hBN but larger than values of most transition metal-dichalcogenide [49]. A negative Poisson's ratio is found in $Si_9C_{15}$ siligraphene (see Fig. 2c), which means that the monolayer $Si_9C_{15}$ intrinsically exhibits an auxetic behaviour due to its unique nonplanar configuration. The mechanism of the auxetic behaviour will be discussed latter. Moreover, in contrast to the isotropy observed in its Young's modulus and shear modulus, the Poisson's ratio of $Si_9C_{15}$ siligraphenes is significantly anisotropic, since a nonmonotonic change is observed in $\nu$ of $Si_9C_{15}$ siligraphenes when $\theta$ changes from $0^o$ to $90^o$. For instance, $\nu$ is found to grow from its maximum value of -0.11 to its minimum value of -0.08 when $\theta$ changes from $0^o$ to $45^o$. When $\theta$ varies from $45^o$ to $90^o$, $\nu$ increases to -0.11. In general, the ratio of the maximum Poisson's ratio to its minimum counterpart can be up to 1.38.

Monolayer $Si_9C_{15}$ was reported to possess a moderate bandgap [37]. To further broaden the application fields of $Si_9C_{15}$ siligraphene in 2D optoelectronic devices, some methods are desired to be developed for modulating its electronic properties. Strain engineering is one of the most common means to modify electronic structures of 2D materials [38-41]. Thus, we also studied the strain effect on the electronic band structures of $Si_9C_{15}$ siligraphene. It is shown in Fig. 3a that $Si_9C_{15}$ siligraphene without loading possesses a direct bandgap of 2.07 eV, which is close to the value of 1.9 eV reported in the previous study [37]. When the size of $Si_9C_{15}$ siligraphene is elongated by 5% or a tensile strain of 5% is applied, the bandgap reduces to 1.55 eV as shown in Fig. 3b. The significant reduction observed in the bandgap of the stretched $Si_9C_{15}$ siligraphene is ascribed to the splitting of both valence band maximum and conduction band maximum near the $\Gamma$ point, though the key features such as the direct band gap of band structures now remain almost unchanged. As the tensile strain increases to 10%, the direct bandgap shifts to an indirect bandgap as shown in Fig. 3c, which is accompanied with a reduction of the bandgap of $Si_9C_{15}$ siligraphene further to 1.48 eV.

### 3.2 Mechanical properties at the finite temperature from MD

The mechanical properties of $Si_9C_{15}$ siligraphene were also investigated by MD simulations, since the MD method has the capability to consider a $Si_9C_{15}$ siligraphene sample with a larger number of atoms. Moreover, in MD simulations, dynamic factors such as the temperature effect can be taken into account in the study of tensile test, which is out of the reach of the above DFT calculations.

In Fig. 4a, we show the relationship between the stress $\sigma$ and strain $\varepsilon$ of $Si_9C_{15}$ siligraphene stretched in the *x* or ZZ direction. Here, the temperature is 10 K. It is found that the stress in $Si_9C_{15}$ siligraphene monotonously grows as the strain increases before its fracture. Moreover, some changes in the slopes of stress-strain curves are observed in this process, which indicate the possible structural changes of $Si_9C_{15}$ siligraphene during the stretching process. To prove this deduction, we also show in Fig. 4a the relationship between the strains in the loading and the corresponding perpendicular directions, i.e., $\varepsilon_x$ and $\varepsilon_y$, respectively. It is found that when the tensile strain $\varepsilon_x$ is smaller than 12%, $\varepsilon_y$ with a positive magnitude increases as $\varepsilon_x$ grows, which indicates the existence of a negative Poisson's ratio in $Si_9C_{15}$ siligraphene during this initial tension process. This phenomenon is consistent with the intrinsic auxetic behaviour observed in above DFT calculations. To explain the intrinsic auxetic behaviour in $Si_9C_{15}$ siligraphene, we show some representative snapshots for the configuration of $Si_9C_{15}$ siligraphene during the tension process in Fig. 5a. It is found that the rippling amplitude of $Si_9C_{15}$ siligraphene decreases markedly as the tensile strain increases, which can result in the expansion of the structure in the perpendicular direction as shown in Fig. 5b. In other words, the de-wrinkling behaviour leads to the negative Poisson's ratio in intrinsically rippled $Si_9C_{15}$ siligraphene. The similar de-wrinkling mechanism was previously used to design the auxetic property of graphene by artificially introducing the topological defects and oxidation [50, 51]. When the tensile strain $\varepsilon_x$ increases beyond 12%, $\varepsilon_y$ shown in Fig. 4a is found to inversely decrease as $\varepsilon_x$ grows, which indicates that the Poisson's ratio of $Si_9C_{15}$ siligraphene becomes positive now. The occurrence of the positive Poisson's ratio demonstrates the disappearance of de-wrinkling effect in the nearly flat $Si_9C_{15}$ siligraphene now.

After the tensile loading shifts from the *x* (or ZZ) direction to the *y* (or AC) direction, although a similar stress-strain relation retains in $Si_9C_{15}$ siligraphene, a different relationship between the strains in the loading direction ($\varepsilon_y$) and the corresponding perpendicular direction ($\varepsilon_x$) is observed. As shown in Fig. 4b, when $\varepsilon_y$ is smaller than 3.3%, $\varepsilon_x$ grows as $\varepsilon_y$ increases due to the similar de-wrinkling effect in the intrinsically rippled $Si_9C_{15}$ siligraphene. When $\varepsilon_y$ reaches 3.3%, an abrupt drop is observed in $\varepsilon_x$, which is induced by a sudden configuration transition from the irregularly rippled shape to a regularly wrinkled shape as shown in Fig. 6. The end of the configuration transition process is observed at the strain of 7.2%. After that, $\varepsilon_x$ increases again as $\varepsilon_y$ grows similarly due to the de-wrinkling of the regularly wrinkled Si$_9$C$_{15}$ siligraphene. When $\varepsilon_y$ becomes larger than 19%, $\varepsilon_x$ turns to decrease with growing $\varepsilon_y$, indicating the flattening of the stretched Si$_9$C$_{15}$ siligraphene as shown in Fig. 6.

From the above discussion, it is found that the deformation of Si$_9$C$_{15}$ siligraphene stretched along the *x* (or ZZ) direction consists of two processes that successively are nonplanar and planar deformations. These two processes are noted here as process I and II, respectively. As for the siligraphene stretched along the *y* (or AC) direction, there exist three processes including the initial nonplanar deformation within the irregularly rippled shape, the successive nonplanar deformation within the regularly wrinkled shape, and final planar deformation, which are noted as process I, II, and III, respectively. The Young's modulus $E$ and the Poisson's ratio $\nu$ at these different processes of Si$_9$C$_{15}$ siligraphene are listed in Tab. 1. Here, the Young's modulus $E$ was calculated as the slope of $\sigma$-$\varepsilon$ curve at the beginning of each process, while the Poisson's ratio $\nu$ was represented as the slope of $\varepsilon_x$-$\varepsilon_y$ (or $\varepsilon_y$-$\varepsilon_x$) curve. The Young's moduli at process I of ZZ and AC directions are around 268 GPa and 247 GPa, respectively. The similarity of Young's moduli in these two directions is consistent with the almost isotropy of the intrinsic Young's modulus observed in above DFT calculations. However, the magnitude of the Young's modulus obtained from MD simulations is generally larger than the result extracted from DFT calculations. This quantitative difference is probably attributed to the different configurations of their intrinsic ripples. Because a larger model and the temperature effect can be considered in MD simulations, Si$_9$C$_{15}$ siligraphene in MD simulations intrinsically exhibits an irregularly rippled shape (see Figs. 5a and 6), while a regularly rippled shape of Si$_9$C$_{15}$ siligraphene is observed in DFT calculations (see Fig. 1b). This difference also results in the different Poisson's ratios obtained from these two methods, though both methods predict that Si$_9$C$_{15}$ siligraphene intrinsically has a negative

Poisson's ratio. Specifically, the Poisson's ratio at process I of ZZ direction is -0.38, which is about three times larger than the value (-0.12) of AC direction. When the $Si_9C_{15}$ siligraphene stretched in the $y$ (or AC) direction enters process II, its Young's modulus increases to 470 GPa, since the deformation in the loading direction of the regularly wrinkled structure shown in Fig. 6 now is majorly due to the elongation or rotation of strong C-C or Si-C bonds rather than the weak de-wrinkling effect. Meanwhile, the magnitude of the negative Poisson's ratio in this process can grow to -1.15, indicating that the de-wrinkling effect now has a more significant effect on the deformation perpendicular to the loading direction. In the final planar deformation process, Young's moduli of ZZ and AC directions of $Si_9C_{15}$ siligraphene are, respectively, 487 GPa and 385 GPa, while Poisson's ratios of these two directions are 0.119 and 0.048, respectively. This anisotropic mechanical property observed in MD simulations is in contrast to the isotropic mechanical property observed in DFT calculations of planar $Si_9C_{15}$ siligraphene (see Fig. 2). The anisotropy of mechanical property is ascribed to the very different states of stress of the planar $Si_9C_{15}$ siligraphene stretched in two different directions. Moreover, due to the finite strain effect, the Young's modulus and Poisson's ratio extracted from MD simulations both are smaller than the results predicted from DFT calculations. From the above discussion, we can come to the conclusion that the mechanical properties of $Si_9C_{15}$ siligraphene can be affected by the finite size, temperature and strain effects, all of which can be considered in MD simulations but are out of the reach of DFT calculations.

We also examined the mechanical responses of $Si_9C_{15}$ siligraphene stretched at different temperatures ranging from 10 K to 500 K. As shown in Fig. 7, $\sigma$-$\varepsilon$ and $\varepsilon_x$-$\varepsilon_y$ (or $\varepsilon_y$-$\varepsilon_x$) relations of all $Si_9C_{15}$ siligraphene structures are similar to each other, in spite of some differences found in their curve slopes, indicating changes in corresponding mechanical properties (Young's modulus and/or the Poisson's ratio) of $Si_9C_{15}$ siligraphene. As an example, in Fig. 8a and 8b we show Young's moduli and Poisson's ratios of intrinsically rippled $Si_9C_{15}$ siligraphene. Indeed, when the temperature increases from 10 K to 500 K, Young's moduli in ZZ and AC directions, respectively, decrease from 268 GPa to 252 GPa and from 247 GPa to 162 GPa. The reduction in the Young's modulus with growing temperate is due to the so-called thermally induced softening effect widely observed in many other 2D materials [40, 52, 53], which is found to be more significant in the AC direction. The Poisson's ratio, however, is found to be almost independent of the temperature. For instance, Poisson's ratios of ZZ and AC directions rise and fall around mean values of -0.35 and -0.11, respectively. The deviations of the average values of ZZ and AC directions from their maximum or minimum values are less than 11% and 12%, respectively.

We also see from Figs. 4 and 7 that, the stress of all $Si_9C_{15}$ siligraphene structures suddenly drops to zero after the strain reaches the threshold value, i.e., the fracture strain. The threshold value of the stress at the fracture strain is termed as the tensile strength. The sudden drop of the stress from the tensile strength to zero reflects the physical process from the nucleation of the initial void to the final rupture induced by the breaking of Si-C bonds (see the Supplementary Material), since, as shown above, the stiffness of Si-C bonds is weaker than that of their C-C counterparts. No significant increment of strain in this process indicates a brittle fracture of $Si_9C_{15}$ siligraphene, irrespective of the temperature or loading direction. As shown in Fig. 8c and 8d, the tensile strength and fracture strain in both ZZ and AC directions similarly decrease as the temperature grows, since the Si-C bonds in $Si_9C_{15}$ siligraphene exhibit a more significant thermal fluctuation at a higher temperature. Once some Si-C bonds are broken due the serious thermal fluctuation at a high temperature, some voids are initiated, which is immediately followed by the catastrophic failure or the brittle facture of whole $Si_9C_{15}$ siligraphene. In terms of the temperature effect on fracture properties in different directions, the tensile strength and fracture strain in the ZZ direction are more sensitive to the temperature change. For instance, when the temperature is as low as 10 K, the tensile strength in the ZZ direction is 168 GPa, which is about 24% larger than the value of 136 GPa in the AC direction. However, due to its more sensitivity to the temperature change, the tensile strength in the ZZ direction decreases to 99 GPa at 400 K, which is comparable to the value of 97 GPa in the AC direction at the same temperature. Similarly, although at the temperature of 10 K the fracture strain in the ZZ direction is similarly about 24% larger than the value of the AC direction, fracture strains in these two directions become close to each other at the temperature of 400 K. As the temperature keeps growing to 500 K, both tensile strength and fracture strain of the AC direction inversely turn to be much larger than their counterparts of the ZZ direction. Moreover, the facture strain is comparable to the value of other stretchable 2D materials such as graphene, h-BN and $MoS_2$ [40].

## 4. Conclusions

The mechanical properties of recently synthesized $Si_9C_{15}$ siligraphene are investigated in this work based on atomistic simulations including DFT calculations and MD simulations. Both methods prove the existence of the intrinsic negative Poisson's ratio in $Si_9C_{15}$ siligraphene, in spite of the difference observed in magnitudes of the negative Poisson's ratios predicted from these two methods. This quantitative deviation is ascribed to the different rippling configurations of $Si_9C_{15}$ siligraphenes predicted from two methods, since MD simulations have the capability to consider the realistic finite size and temperature effects, which are, however, out of the reach of DFT calculations. The de-wrinkling of intrinsic rippled configurations of $Si_9C_{15}$ siligraphene under uniaxial tension induces the auxetic behaviour or the negative Poisson's ratio of $Si_9C_{15}$ siligraphene. Moreover, different de-wrinkling behaviours are observed in $Si_9C_{15}$ siligraphene stretched along different directions. A continuous de-wrinkling process of irregular ripples is observed in $Si_9C_{15}$ siligraphene stretched along the ZZ direction, while a transition from the irregularly rippled shape to the regularly wrinkled shape is observed in the de-wrinkling process of $Si_9C_{15}$ siligraphene stretched along the AC direction. The configuration transition can greatly enhance the de-wrinkling effect and thus increase the magnitude of the negative Poisson's ratio in the AC direction. In addition, $Si_9C_{15}$ siligraphene structures stretched in ZZ and AC directions also have different fracture properties, but both of them similarly possess relatively large fracture strains. The stretchability of $Si_9C_{15}$ siligraphene together with its strain-sensitive bandgap observed in DFT calculations indicates the feasibility of strain engineering in modulating the electronic properties of $Si_9C_{15}$ siligraphene. With the combination of unique auxetic property, excellent mechanical property and tunable electronic property, $Si_9C_{15}$ siligraphene may have new potential applications of in nanodevices and nanomaterials.

**Acknowledgments**

This work was supported by the Guangdong Basic and Applied Basic Research Foundation (Grant No. 2022A1515010631) and Shenzhen Science and Technology Program (Grant No. GXWD20220811164345003).

**Figures**

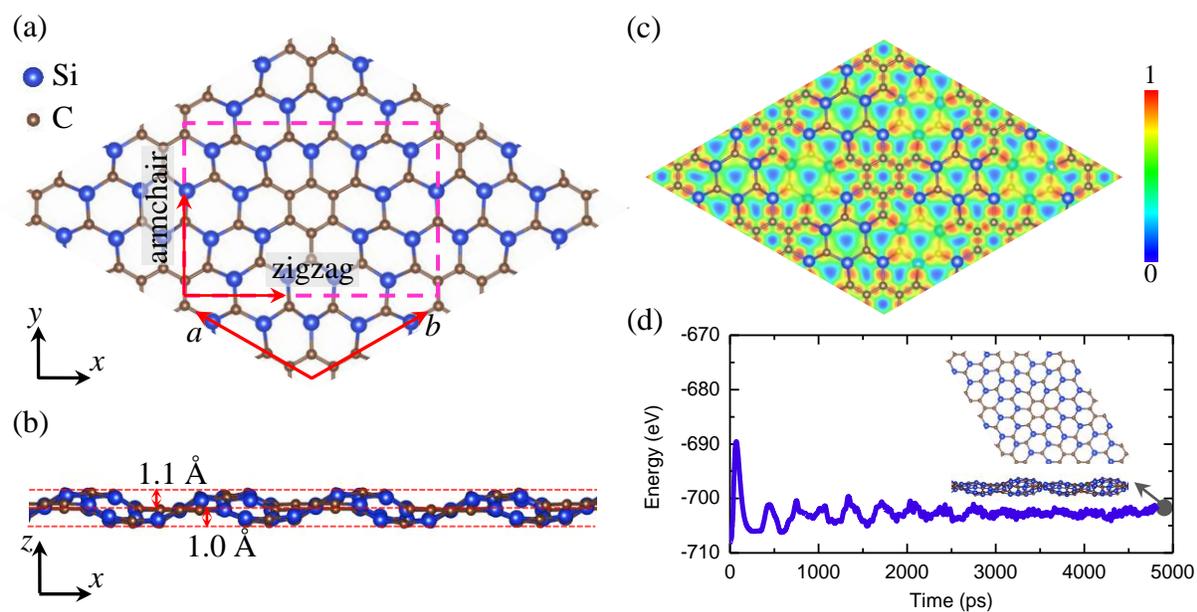

**Fig. 1.** (a) Top view and (b) side view of the monolayer $Si_9C_{15}$ after the structural optimization in DFT calculations. (c) The ELF of the optimized monolayer $Si_9C_{15}$. (d) Energy evolution of monolayer $Si_9C_{15}$ during AIMD simulations at the temperature of 500 K. The inset shows the structure extracted at the end of AIMD simulations.

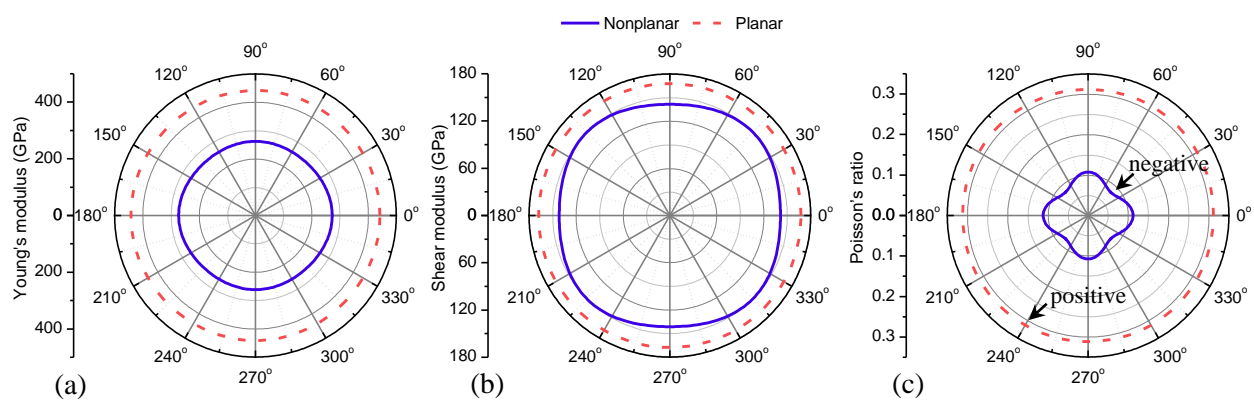

**Fig. 2.** Orientation-dependent (a) Young's modulus, (b) shear modulus and (c) Poisson's ratio of original nonplanar and planar $Si_9C_{15}$ siligraphene.

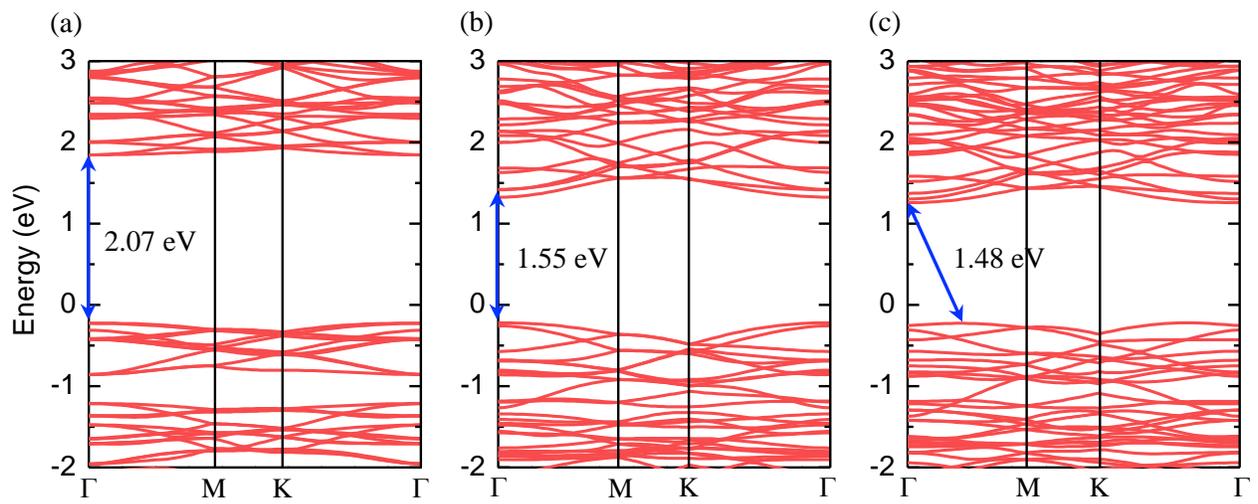

**Fig. 3.** Band structures of monolayer $Si_9C_{15}$ (a) without strain and with strains of (b) 5% and (c) 10%.

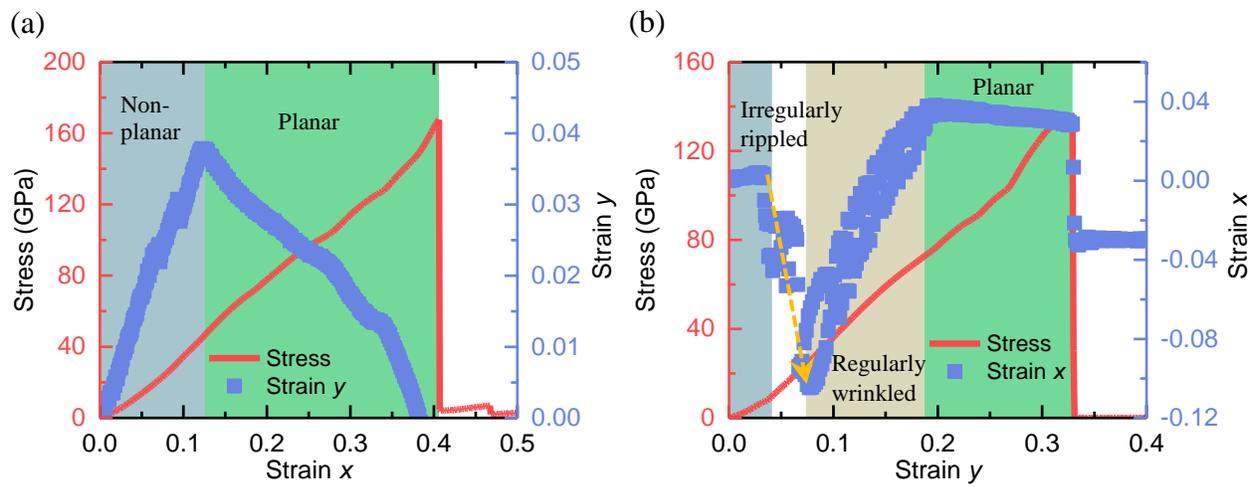

**Fig. 4.** The stress-strain relationship and the relationship between transverse strain and applied axial strain of $Si_9C_{15}$ siligraphene stretched in (a) the *x* (or ZZ) direction and (b) the *y* (or AC) direction.

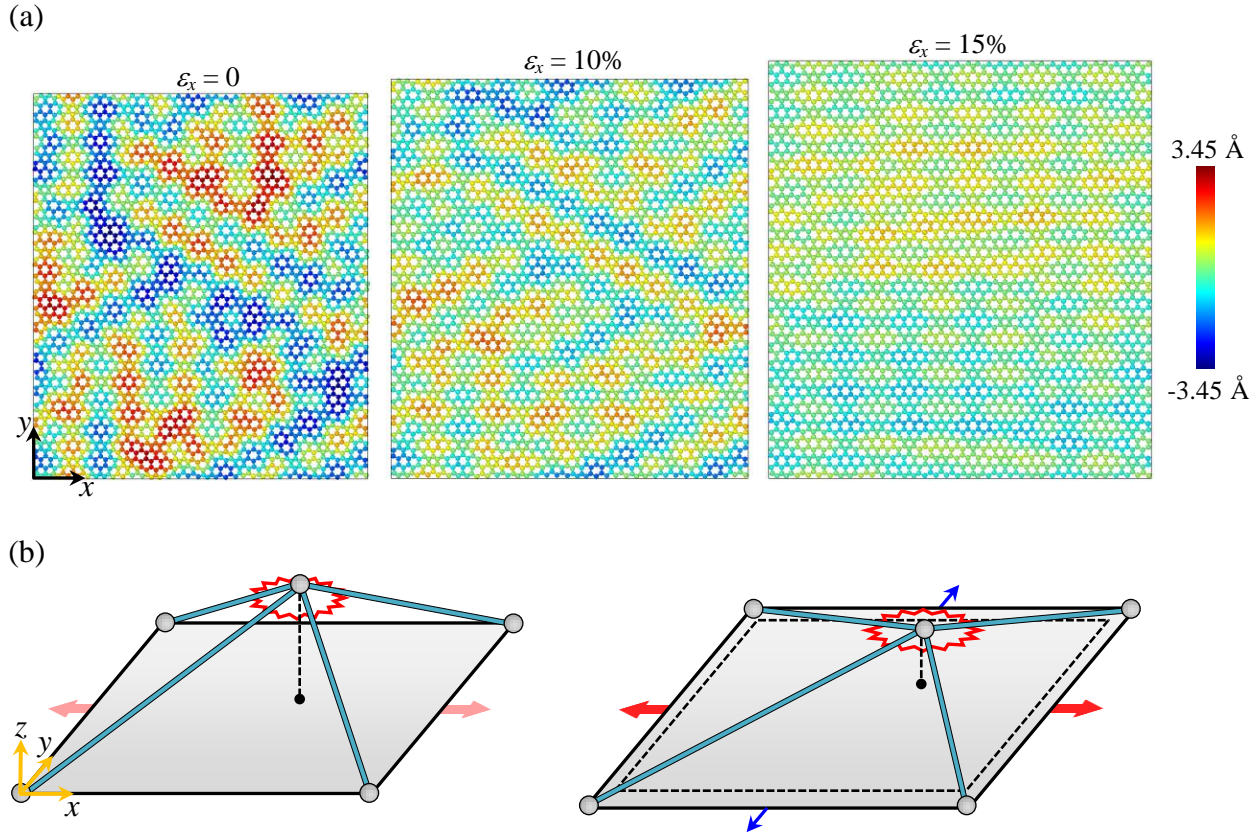

**Fig. 5.** (a) The initial configuration ($\varepsilon_x = 0$) of Si$_9$C$_{15}$ siligraphene and deformed configurations at different uniaxial tension strains in the $x$ direction. Here, contours illustrate the out-of-plane deformation of the siligraphene. (b) A stick-spirals model of the Si$_9$C$_{15}$ siligraphene at the initial configuration (left) and deformed configuration (right) used to demonstrate the mechanism of the intrinsic auxetic behaviour of Si$_9$C$_{15}$ siligraphene.

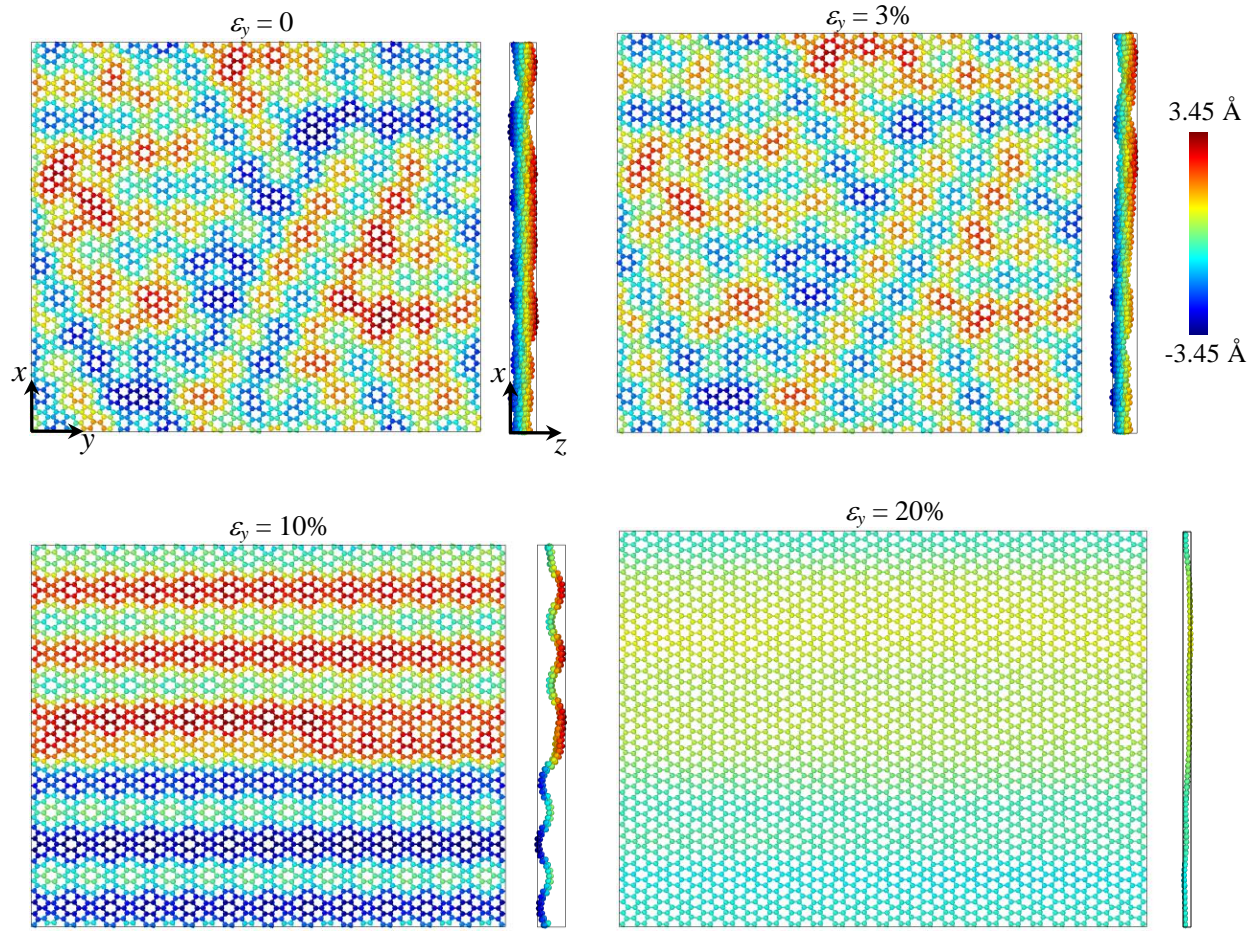

**Fig. 6.** (a) The initial configuration ($\varepsilon_y = 0$) of Si$_9$C$_{15}$ siligraphene and deformed configurations at different uniaxial tension strains in the *y* direction. Here, contours illustrate the out-of-plane deformation of the siligraphene.

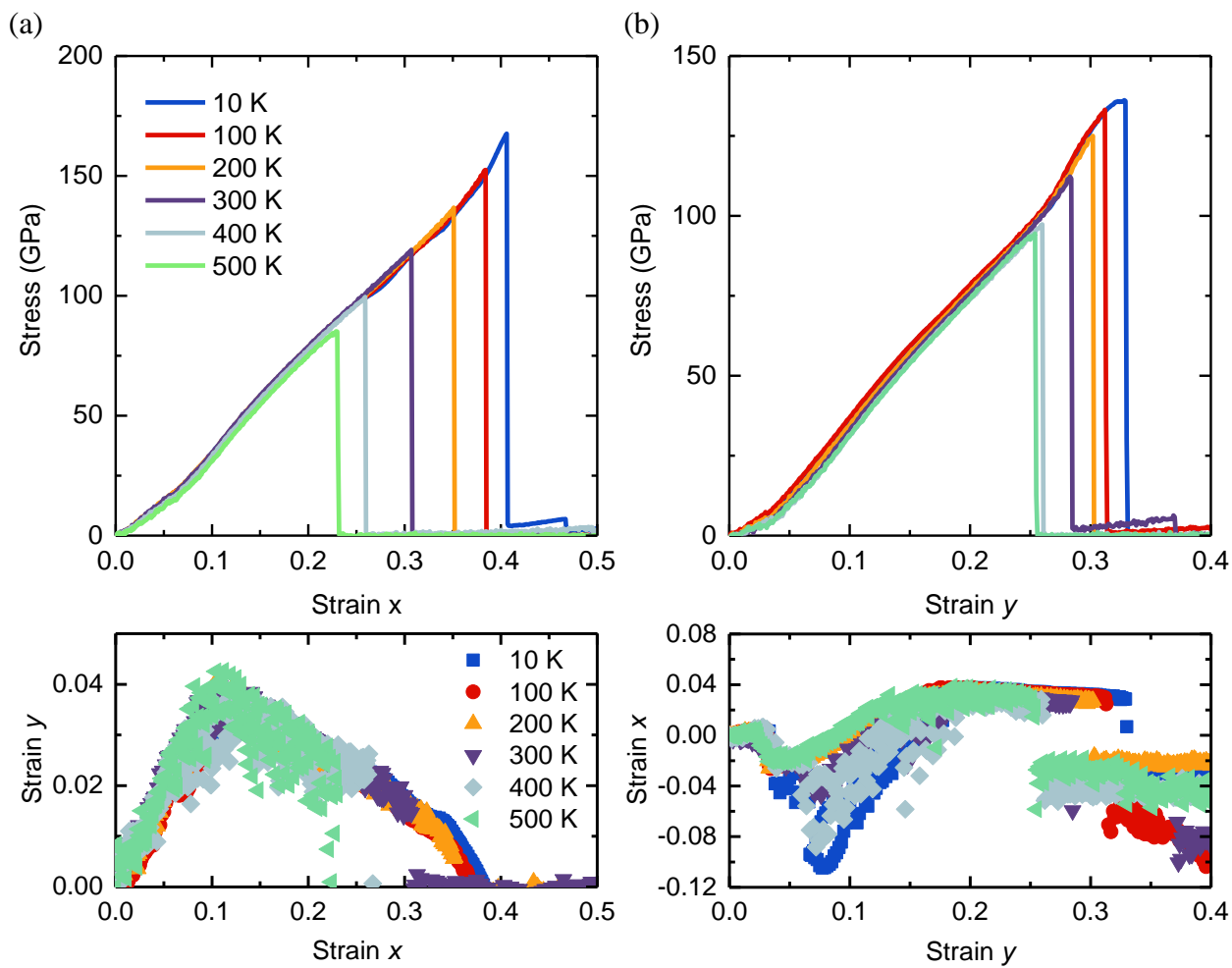

**Fig. 7.** The stress-strain relationship (top panel) and the relationship between transverse strain and applied axial strain (bottom panel) of Si$_9$C$_{15}$ siligraphene stretched in (a) *x* and (b) *y* directions at different temperatures.

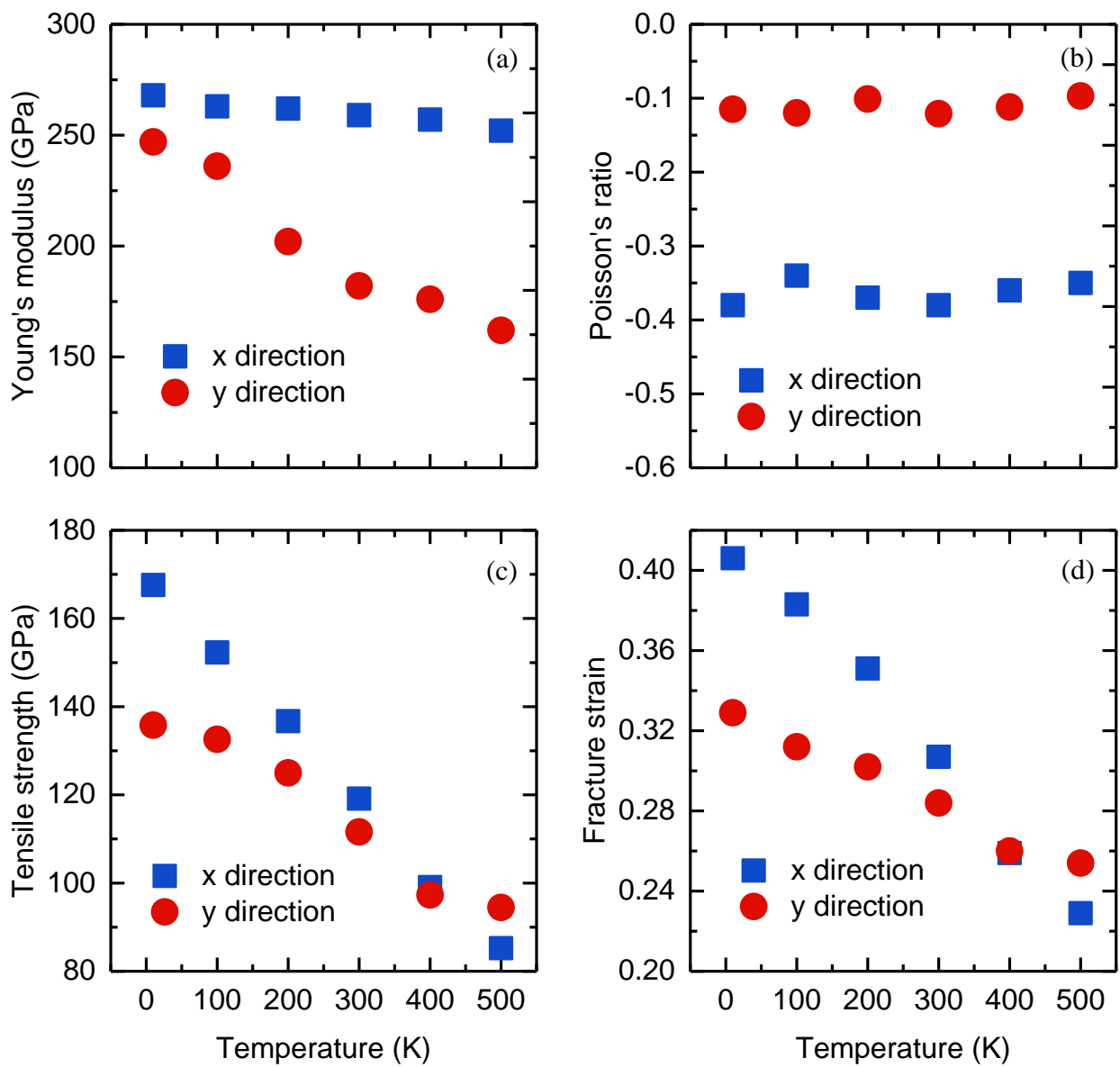

**Fig. 8.** (a) Young's modulus, (b) Poisson's ratio, (c) tensile strength, and (d) fracture strain of the Si$_9$C$_{15}$ siligraphene at different temperatures.

**Table**

Tab. 1. The intrinsic Young's modulus and Poisson's ratio of $Si_9C_{15}$ siligraphene at different deformation processes during the uniaxial tension

|  | ZZ direction | | AC direction | | |
| --- | --- | --- | --- | --- | --- |
|  | I | II | I | II | III |
| Young's modulus (GPa) | 268 | 487 | 247 | 470 | 385 |
| Poisson's ratio | -0.38 | 0.12 | -0.12 | -1.15 | 0.05 |